\documentclass[twocolumn,superscriptaddress]{revtex4-2}
\usepackage{gnuplottex}
\usepackage{mathtools}
\usepackage{float}
\usepackage{array}
\usepackage[english]{babel}
\usepackage[utf8]{inputenc}
\usepackage{amsmath}
\usepackage[table]{xcolor}
\usepackage{amsfonts}
\usepackage{graphicx}
\usepackage{dsfont}
\usepackage[colorlinks=true, linkcolor=blue, citecolor=dred]{hyperref}
\usepackage{algorithm}
\usepackage[noend]{algpseudocode}
\usepackage{bbold}
\usepackage{mathtools}
\usepackage{amssymb}
\usepackage{bm}
\usepackage{bbm}
\usepackage{tabularx, booktabs}
\usepackage{xspace}
\usepackage{listings}
\usepackage[noend]{algpseudocode}
\newcolumntype{Y}{>{\centering\arraybackslash}X}
\definecolor{dgreen}{rgb}{0,.5,0}
\definecolor{dblue}{rgb}{0,0,.5}
\definecolor{dred}{rgb}{0.5,0,.5}

\usepackage{physics}

\newcommand\reallywidehat[1]{%
\savestack{\tmpbox}{\stretchto{%
  \scaleto{%
    \scalerel*[\widthof{\ensuremath{#1}}]{\kern-.6pt\bigwedge\kern-.6pt}%
    {\rule[-\textheight/2]{1ex}{\textheight}}
  }{\textheight}%
}{0.5ex}}%
\stackon[1pt]{#1}{\tmpbox}%
}
\begin{document}


\title{Defect-Mediated Pairing and Dissociation of Strongly Correlated Electrons\linebreak in Low Dimensional Lattices: \textit{The Quantum Taxi Effect} }

\author{Vincent Pouthier}
\affiliation{Universit\'{e} Marie et Louis Pasteur, CNRS, Institut UTINAM (UMR 6213), Equipe $\phi$th, F-25000 Besan\c {c}on, France} 
\author{Saad Yalouz} 
\email{saad.yalouz@cnrs.fr}
\affiliation{Laboratoire de Chimie Quantique de Strasbourg, Institut de Chimie,
CNRS/Université de Strasbourg, 4 rue Blaise Pascal, 67000 Strasbourg, France}

\begin{abstract}

We study the quantum dynamics of a strongly correlated electron pair in a one-dimensional lattice, focusing on the occurrence of local dissociation/pairing mechanisms induced by a site energy defect. \linebreak To this end, we simulate the time evolution of two interacting electrons on a finite-size chain governed by an extended Hubbard Hamiltonian including on-site Coulomb repulsion \( U \) and nearest-neighbor interaction $V$, along with single-electron hopping $J$. By introducing a local site energy defect with amplitude \( \Delta \), we show that a transition between spatially paired/dissociated electrons can occur in the vicinity of this site. Such mechanisms arise in a strongly correlated regime with non-zero nearest neighbor Coulomb interactions and under the conditions  $ (U \sim V \sim \Delta) \gg J$. 
To rationalize these phenomena, we reformulate the two-electron dynamics of the original Hubbard chain as an effective single-particle problem on a two-dimensional network. 
Within this framework, we show that the pairing/dissociation dynamics are driven by resonances between two distinct families of two-electron eigenstates:  \textit{(i)} states with two spatially well-separated electrons with one located at the site defect, and \textit{(ii)} states with locally bound electron located away from the defect.
At resonance, these states hybridize, allowing transitions from locally paired to dissociated electrons (and vice versa) in the vicinity of the defect. These results provide new insights into exotic pairing phenomena in strongly correlated electronic systems and may have implications for the design of tunable  many-body states in low-dimensional quantum materials.

\end{abstract} 
 
\maketitle

\section{Introduction}

Trying to predict and rationalize the mechanisms that lead to electron pairing and dissociation in matter is one of the most fundamental questions addressed in electronic structure theory.  
From chemical bond theory~\cite{lewis1916atom,van1935quantum, coulson1947representation,coulson1949xxxiv,atkins2011molecular} to the description of charge transport in crystal structures~\cite{bloch1929quantenmechanik,wilson1931theory,slater1954simplified,bouckaert1936theory,ashcroft1976solid}, this long-standing question has always been central to understanding the properties of novel materials and/or molecular systems in both condensed matter physics and quantum chemistry.

From a theoretical standpoint, it is well established that the emergence of pairing and dissociation mechanisms strongly depends on the nature of the electron-electron interaction regime governing the system under consideration. 
In weakly correlated systems, where these interactions are relatively minor, single-particle frameworks (\textit{i.e.} orbital or band theory) often provide sufficiently accurate descriptions of electronic behavior.\linebreak
In contrast, strongly correlated systems, where electron-electron interactions play a dominant role, cannot be adequately described by single-particle approaches. In these cases, the electronic structure becomes significantly more complex due to the intrinsic many-body nature of the system.
This complexity gives rise to unconventional pairing and dissociation pathways which are known to underlie a wealth of emergent quantum phenomena.
Examples, among others, include supraconductivity~\cite{bardeen1957theory, yanase2003theory, fausti2011light}, singlet fission in organic semiconductors~\cite{miyata2019triplet, zimmerman2011mechanism, berkelbach2013microscopic, smith2010singlet}, magnetism in molecular compounds~\cite{roseiro2025interplay, roseiro2023modifications, sheng1994magnetism, verot2012importance, vela2017electron}, and various photoinduced effects~\cite{gao2020photoinduced, kaneko2019photoinduced, zhang2022steady, wan2024proof, shankar1973photon}.
Despite decades of research on the subject, a comprehensive theoretical understanding of electron pairing/dissociation mechanisms in strongly correlated systems still remains an open issue. 
And this question becomes even more complex when one considers the additional influence of a natural (and often unavoidable) feature in materials: the presence of \textit{local defects}.

At the single-electron level, the impact of local defects on the properties of charge (or energy) carriers has been extensively studied in the literature.
A broad range of effects has been identified, ranging from strong localization (\textit{e.g.} Anderson localization~\cite{anderson1958absence}) to enhanced delocalization, depending on the symmetry and structure of the underlying system~\cite{yalouz2022extended,pepe2024optimized,chavez2021disorder,yalouz2020continuous,aizenman2011extended,aizenman2013resonant,pouthier2014disorder}.
However, when the focus shifts to many-electron systems, rationalizing the influence of local defects is far from trivial and still represents an open question (see Refs.~\cite{efros2012electron,vojta2019disorder} and references therein).
In such case, the physical landscape   becomes significantly richer due to the interplay among diverse many-electron configurations (\textit{e.g.} bound pairs, dissociated/localized states \textit{etc.}).
Defects can induce resonant behaviors among these configurations,  
giving rise to exotic states hybridization (with no analogue in the single-particle framework) that will change the pairing and dissociation mechanism of correlated electrons.
A comprehensive understanding of these phenomena is therefore essential if we want to better predict the arising properties of strongly correlated materials with defects.  


\pagebreak

Prompted by these considerations, this work investigates the minimal scenario of a single strongly-correlated electron pair evolving on a one-dimensional (1D) lattice, focusing on local dissociation and pairing mechanisms triggered by the presence of a site energy defect.\linebreak
To this end, we simulate the time evolution of the two-electron system on a finite-size chain described by an extended Hubbard Hamiltonian~\cite{hubbard1963electron}, which includes on-site Coulomb repulsion $U$, nearest-neighbor interaction $V$, and single-electron hopping $J$.
Our goal is to examine how the site defect modifies the stability and propagation  of bound electron pairs, revealing the competition between interaction-driven pairing mechanisms and defect-induced scattering.
By analyzing the time evolution under varying interaction regimes and defect configurations, we evidence a novel dynamical phenomenon that we term ``\emph{Quantum Taxi Effect}'' (QTE).
In this process, an incoming (locally bound) electron pair encountering the defect may dissociate, with one electron becoming localized at the defect while the other moves away. 
Conversely, a free electron approaching a defect already occupied by another electron may resonantly bind to form a pair, which then propagates as a single unit. 
This resonance-driven behavior marks a significant deviation from conventional transport mechanisms and underscores the complex role of electron correlations in defect-laden quantum systems. 
Our findings might open new directions for exploring defect-assisted transport in correlated systems with potential applications in quantum information processing and quantum device engineering.

The paper is organized as follows. In Sec.~\ref{sec:II}, the extended Hubbard Hamiltonian describing the electron properties in a molecular chain is introduced. 
Then, the two-electron dynamics in the singlet subspace is presented through its equivalence with that of a single particle moving on a 2D lattice. 
In Sec.~\ref{sec:III}, the simulation of the Schrodinger equation is performed and the numerical results
are presented. 
These results are discussed and interpreted in Sec.~\ref{sec:IV}. 
Finally, we summarize our findings and discuss future directions in Sec.~\ref{sec:V}.

\begin{figure*}
    \centering
    \includegraphics[width=17cm]{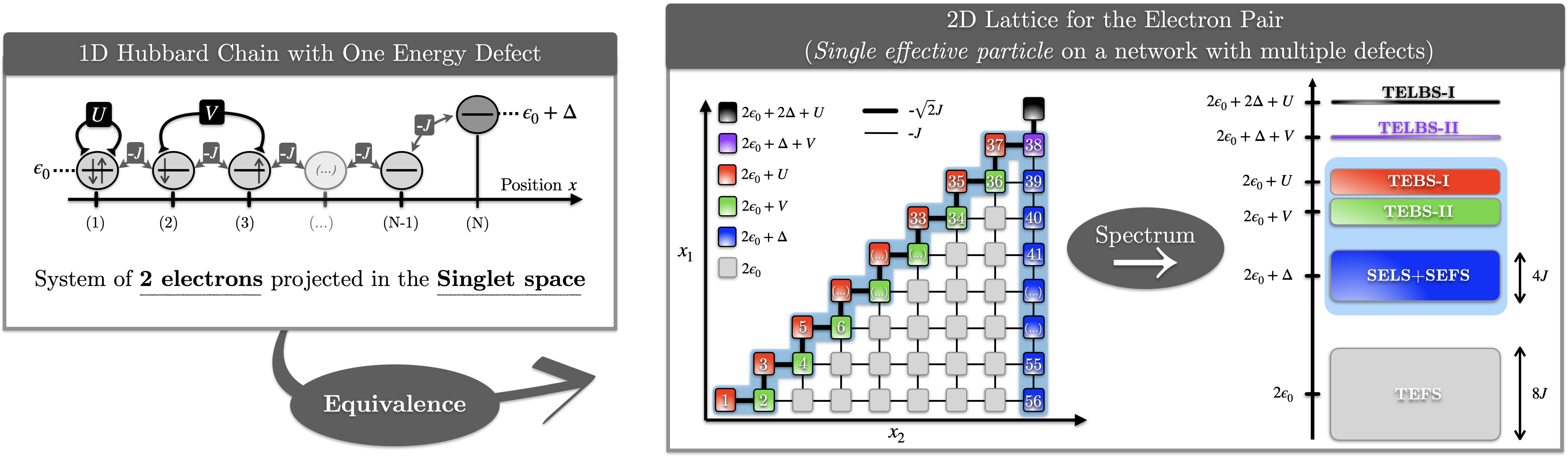}
    \caption{{Equivalence between the 1D Hubbard chain and a 2D lattice in the case of the two-electron dynamics.}\linebreak
\textit{{Left panel:}} Illustration of the original system, consisting of a finite 1D Hubbard chain with an energy defect at site $N$. \linebreak
We consider the dynamics of two electrons in the singlet subspace, where both on-site and nearest-neighbor Coulomb interactions ($U$ and $V$ terms in Eq.~(\ref{eq:Hubbard})) are taken into account.
\textit{{Right panel:}} Representation of the equivalent 2D lattice, where the dynamics of the electron pair is mapped onto those of a single effective particle described by a tight-binding model with various types of local energy defects.
Each site in the 2D lattice corresponds to a CSF, as defined in Eq.~(\ref{eq:basis}). 
The group of effective sites (\textit{i.e.}, CSFs) highlighted in light blue marks the active singlet states that define the reaction pathway for the dissociation/paring of electron through occupation of the site carrying the energy defect $\Delta$ on the original system (see left panel).
On the right, the energy spectrum of the 2D lattice is displayed, showing distinct subspaces that correspond to CSFs with different spatial characteristics (such as electronic localized or extended states, as discussed in the main text).
The light blue region again highlights the active singlet subspace forming the relevant reaction pathway for electron pairing/dissociation.
} 
\label{fig:lattice}
\end{figure*}

\section{Theoretical background} 
\label{sec:II}
To elucidate how a local defect can influence electron pairing and dissociation mechanisms in a strongly correlated medium, we consider in this work the minimal scenario of a single electron pair evolving on a 1D extended Hubbard chain with one local site defect.
As illustrated in Figure~\ref{fig:lattice} (see left panel), the chain consists of $N$ sites, labeled by the index $x=1,…,N $. 
We assume that each site $x$ hosts a local orbital (\textit{e.g.} Wannier-like) noted $\varphi_x$ with energy $\epsilon_x$.
All sites are identical, with the same orbital energy $\epsilon_N=\epsilon_0$, except that of the right extremity of the chain $x=N$ which carries an energy defect (of amplitude $\Delta$) such that $\epsilon_N=\epsilon_0+\Delta$. 
An electron can hop between the site $x$ 
to its nearest neighbors $x\pm 1$
with a hopping constant $-J$.
In this lattice, the electrons do not propagate freely but are correlated due to the repulsive Coulomb interaction.
Following the extended Hubbard-model~\cite{hubbard1963electron}, we consider the local repulsive interaction $U$ between two electrons with opposite spins occupying the same site, and $V$ the non-local repulsive interaction between two electrons occupying two nearest-neighbor sites. 
Within these definitions, the electron dynamics is described by the extended Hubbard Hamiltonian expressed as 
%
\begin{equation}\label{eq:Hubbard}
\begin{split}
H&= \sum_{x} \epsilon_x n_x - \sum_{\langle x,x' \rangle } J \sum_{\sigma \in \lbrace \uparrow , \downarrow \rbrace}   c^{\dag}_{x'\sigma} c_{x\sigma} 
\\
&+\sum_{x}  U n_{x\uparrow}n_{x\downarrow}+\sum_{\langle x, x' \rangle}  V n_x n_{x'} ,
\end{split}
\end{equation}
where $c^{\dag}_{x\sigma}$ and $c_{x\sigma}$ are the standard fermion operators that create or annihilate an electron of spin $\sigma$ in the orbital (\textit{i.e.} site) $\varphi_x$, 
$n_{x\sigma}= c^{\dag}_{x\sigma}c_{x\sigma}$ is the number operator that counts the electron number of spin $\sigma$ in the orbital $\varphi_x$ and where $n_x=n_{x\uparrow}+n_{x\downarrow}$. The notation ${\langle x, x' \rangle}$ restrict the summation only on direct nearest-neighbors (without double counting).



\subsection{1D Hubbard Chain Carrying Two Electrons: \linebreak Equivalence with a Single Particle on a 2D Lattice}

Because we want to study the interplay between on-site correlated bound-states and spatially separated electron pairs, special attention will be paid to the singlet subspace.  
In this subspace, the two electrons can occupy a same site and experience both local (on-site $U$) and non-local (inter-site $V$) repulsive Coulomb interaction. 
To generate the entire singlet subspace, we introduce the number state basis (\textit{i.e.} Slater determinants) as follows:
for $x_1=1,...,N$ and $x_2=1,...,N$ we define $N^2$ orthogonal number states $\ket{x_1,x_2}= c^{\dag}_{x_1\uparrow}c_{x_2\downarrow}^\dagger| \varnothing\rangle$ with the arbitrary convention that spin up is to the left while spin down is to the right. 
This convention avoids double counting of equivalent configurations due to the anti-commutation relations satisfied by the fermion operators.
However, as we will specifically focus on singlet electron pairs (with total spin $\langle \hat{S}^2 \rangle  = 0$), it happens that considering the full Slater determinant basis $\{ \ket{x_1,x_2} \}$ is unnecessary because it includes both singlet and triplet electronic states with a vanishing $\hat{S}_z$ component of the total spin.
To build the singlet subspace, we need to symmetrize the basis building the so-called Configuration-State-Functions (CSF), which are simultaneously eigenstates of both spin operators $\hat{S}_z$ and $\hat{S}^2$~\cite{helgaker2013molecular}. 
The two-electron singlet space is described by a total of $N(N+1)/2$ singlet CSFs defined as
%
%
\begin{equation}\label{eq:basis}
 | { \Phi_{x_1,x_2}^\text{Singlet}}\rangle 
= \left\{
    \begin{array}{ll}
          | x_1, x_2 \rangle & \mbox{if } x_2=x_1. \\
    \frac{1}{\sqrt{2}}(  | x_1, x_2 \rangle + | x_2, x_1 \rangle     )      & \mbox{if} \  x_2>x_1.
    \end{array}
\right. 
\end{equation} 
Projecting the extended Hubbard Hamiltonian (Eq.~(\ref{eq:Hubbard})) in this singlet basis, we obtain a Configuration-Interaction Hamiltonian whose shape reads 
\begin{equation}\label{eq:Ham_singlet}
    H^\text{Singlet} = \sum_{(x_1,x_2)} \sum_{(x_1',x_2')} H_{(x_1,x_2),(x_1',x_2')}^\text{Singlet}\ketbra{\Phi_{x_1,x_2}^\text{Singlet}}{\Phi_{x_1',x_2'}^\text{Singlet}},
\end{equation} 
where we introduced the shorthand notation for the couple of indices $ (x_1,x_2)$ always fulfilling $x_2\geq x_1$. 

Let us now adopt a new perspective on the two-electron problem by examining the \emph{connectivity graph} defined by the matrix elements of the singlet Hamiltonian, $H_{(x_1,x_2),(x_1',x_2')}^\text{Singlet}.$
As illustrated in the right panel of Fig.~\ref{fig:lattice}, the way the CSFs are connected through this Configuration Interaction Hamiltonian reveals a graph structure resembling a two-dimensional lattice with multiple defects (energy site and hopping).
In this framework, each CSF acts as an effective site, characterized by local \emph{effective} energies that depend on the system parameters $\epsilon_0$, $\Delta$, $U$, and $V$. These sites are connected by hopping amplitudes of two types, either $-J$ or $-\sqrt{2}J$, depending on the nature of neighbor CSFs.
From a practical standpoint, this structure highlights a direct equivalence between the original two-electron quantum dynamics on the 1D Hubbard chain and that of an \emph{effective single particle} propagating on the 2D lattice shown in Fig.~\ref{fig:lattice}. \linebreak
The dynamics of this effective particle is governed by a \emph{tight-binding (-like) Hamiltonian}, featuring on-site energies and nearest-neighbor hopping terms.
Coulomb repulsion and site energy defect from the original 1D Hubbard-chain (\textit{i.e.} $U,V$ and $\Delta$ terms) manifest now as \emph{effective defect sites in the 2D lattice} (see colored squarre sites in right panel of Fig.~\ref{fig:lattice}).
These inhomogeneities determine whether the \textit{effective particle} evolving on this 2D-lattice is localized/delocalized, which directly reflects whether the ``real'' two-electron state is bound, localized, and/or freely extended.

Note that, for the sake of clarity and conciseness, we did not present here the mathematical steps leading to the final form of the equivalent 2D lattice shown in Fig.~\ref{fig:lattice}. 
However, we refer the interested reader to Appendix~\ref{app:Demo} where we provide a detailed exposition of the approach.



\subsection{ Locally Paired/Unpaired Eigenstates: 
\\ Classification from the
2D-lattice Spectrum  }


\begin{figure*}
    \centering
    \includegraphics[width=15cm]{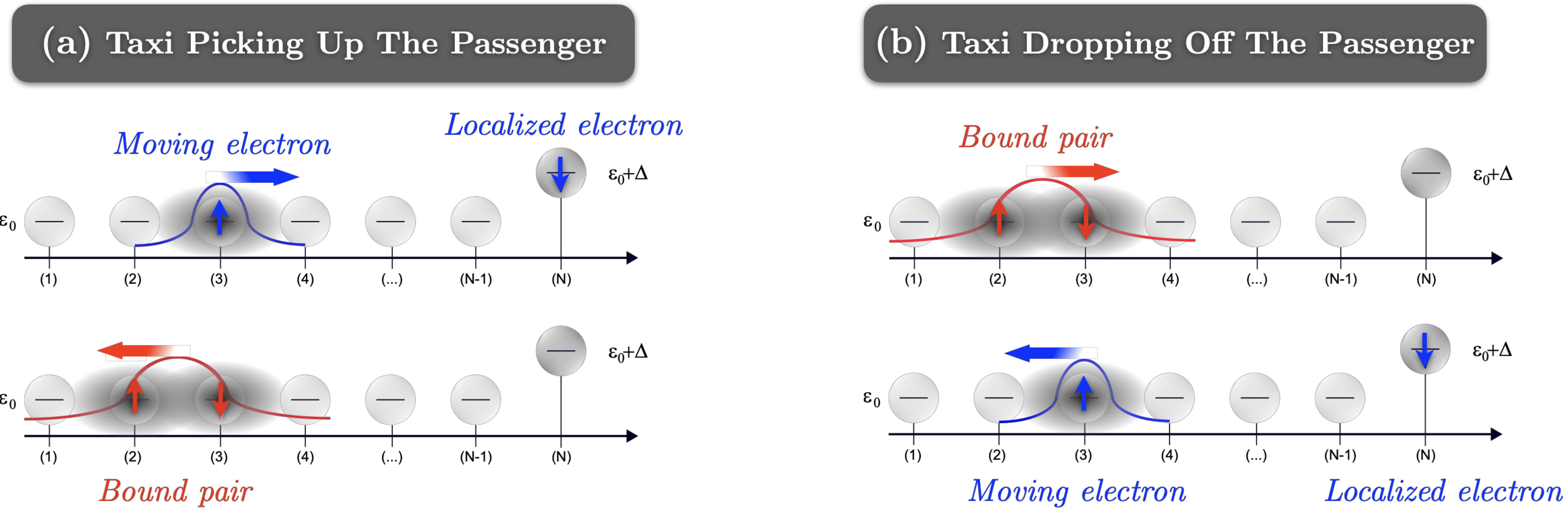}
    \caption{{Illustration of the \textit{Quantum Taxi Effect} (QTE).}  \textbf{(a)} An electron propagating toward the defect occupied by another electron, can form a pair with the localized electron by resonance, the newly formed pair propagating away from the defect.
    \textbf{ (b)} A pair of electrons traveling toward the defect can dissociate, resulting in the localization of one electron on the defect while the other propagates away. 
    }
    \label{fig:illustration}
\end{figure*}

In our study, we will investigate how the two electrons dynamically pair and dissociate over time. 
To this end, we analyze the time evolution of the two-electron wavefunction $\ket{\Psi(t)}$, governed by the time-evolution operator $U(t) = \exp(-i t H^\text{Singlet})$, such that
\begin{equation}
    \ket{\Psi(t)} = U(t) \ket{\Psi(0)},
\end{equation}
where $\ket{\Psi(0)}$ denotes the initial state of the system, to be specified later.
The time evolution is intrinsically connected to the eigenstates $\ket{\psi_{\mu}}$ and eigenenergies $E_{\mu}$ of the Hamiltonian $H^\text{Singlet}$ through the spectral decomposition:
\begin{equation}
    U(t) = \sum_{\mu} e_{\mu}^{-i E_{\mu} t} \ketbra{\psi_{\mu}}{\psi_{\mu}}.
\end{equation}

Understanding the properties of the eigenstates $\lbrace \psi_{\mu} \rbrace $ is therefore essential for rationalizing the mechanisms underlying the pairing and dissociation of the two electrons. 
Although an exact characterization of the eigenstates is challenging due to the complexity of the Hamiltonian, we begin with a qualitative analysis based on the assumption of a well-separated energy spectrum.
This spectral structure is naturally motivated by the topology of the equivalent 2D lattice, which features various types of defects. 
Assuming strongly correlated regime where $(U \sim V \sim \Delta) \gg J$, these inhomogeneities  lead to a non-uniform energy distribution, prompting the spectral resolution approach developed in the following paragraphs.

Based on these observations, in right panel of Fig.~\ref{fig:lattice} we have displayed a qualitative resolution of the spectrum of the 2D-lattice (which is equivalent to the one of the 1D-Hubbard chain in singlet space).
As readily seen here, the system exhibits six different types of eigenstates. 

First, the two-electron free states (TEFS) which form an energy band centered around $2\epsilon_0$ as shown in right panel of Fig.~\ref{fig:lattice}. 
On the 2D lattice, these states will essentially describe the delocalization of the fictious particle on the grey squares sites which form the central part of the network (far away from the defects on the periphery).
On the original 1D-Hubbard chain, such states refer to two quasi-independent electrons freely propagating on the sites which do not carry any energy defect.

Second, the Single-Electron-Localized and Single-Electron-Free states (SELS+SEFS).
In the 2D-lattice, these states will describe the motion of the fictious particle along the line $x_2 = N$ (blue squares), leading to an energy band centered around $2\epsilon_0 + \Delta $ (see right panel of Fig.~\ref{fig:lattice}).
In the 1D-Hubbard chain, these eigenstates correspond to the localization of one electron at the defect site $N$, and the delocalization of the other one on the other sites which do not carry the energy defect.

Third and fourth, there exist two types of Two-Electron-Bound-States (TEBS-I and TEBS-II). 
On the 2D lattice, TEBS-I  will correspond to the localization of the fictitious particle near the row $x_2 = x_1$ (red squares), leading to the formation of an energy band centered around $2\epsilon_0 + U$ (see right panel of Fig.~\ref{fig:lattice}).
On the 1D Hubbard chain, these eigenstates will refer to pair states whose internal structure preferentially involves two electrons trapped at the same site.
On another note, TEBS-II, on the 2D lattice will be associated with the localization of the fictitious particle near the row $x_2 = x_1 + 1$ (green circles), forming then an energy band centered around $2\epsilon_0 + V$.
On the 1D chain, these states will describe two electrons remaining trapped at direct neighboring sites.

%

Finally, the system supports two types of Two-Electron Localized Bound States (TELBS-I and TELBS-II). 
Starting with TELBS-1, on the 2D lattice this state corresponds to the localization of the fictitious particle on the site $x_2 = x_1 = N $ (unique black square), leading to the highest energy in the spectrum approximately equal to $2\epsilon_0 + U + 2\Delta$.
In the 1D Hubbard chain, TELBS-I will describe the unique paired state where the two electrons are simultaneously localized on the defect of the 1D chain (on site $N$). 
In contrast, TELBS-II in the equivalent 2D lattice corresponds to the localization of the fictitious particle near the site $x_2 = x_1 + 1 = N$ (unique square with purple color) with an energy approximately equal to $2\epsilon_0 + V + \Delta$.  
In the 1D Hubbard chain, it defines the unique state in which the electrons are trapped at the neighbor sites $N-1$ and $N$.

\subsection{ Pathway for Dynamical Two-Electron Pairing/Dissociation \textit{via} States Hybridization: \linebreak  \textit{The Quantum Taxi Effect} }

As discussed previously, we see that the 2D lattice shown in Fig.~\ref{fig:lattice} provides a valuable framework for a comprehensive representation of the two-electron energy spectrum (right panel in Fig.~\ref{fig:lattice}). 
This spectrum clearly highlights the possibility of resonances between different types of states, depending on the original model parameters ($U,V,\epsilon_0,\Delta$ and $J$). 
However, depending on the nature of these resonances, a variety of dynamical behaviors can emerge. 


In this paper, we focus specifically on a process involving resonances between TEBS and SELS+SEFS that occur when $U \approx V \approx\Delta$.
Under these conditions, we expect that the hybridization between TEBS and SELS+SEFS promotes the emergence of a dynamical phenomenon we call the \textit{Quantum Taxi Effect} (QTE) illustrated in Fig.~\ref{fig:illustration}.
During this process, a free electron approaching the defect, which is already occupied by another electron, can form a locally bound pair with the localized electron by resonance, with the newly formed pair subsequently propagating away from the defect. Conversely, a locally bound pair of electrons traveling toward the defect can dissociate, resulting in one electron localizing to the defect while the other propagates away, far from the defect.
The next section is thus devoted to the characterization of the QTE through the numerical resolution of the time-dependent Schrodinger equation.

\section{Numerical Simulation}
 \label{sec:III}
 
In our numerical study, we solve the time-dependent two-electron Schrodinger equation on a Hubbard chain containing $N=20$ sites to characterize the arising of dissociation/pairing mechanisms. 
Throughout the study, the hopping constant is taken as the unit of energy, with $J=1$. 
The on-site Coulomb interaction between two electrons with opposite spins is fixed at $U=10 $.
The values of $V$ and $\Delta$ will be precised later on depending on the study case.

Among the various observables that can be extracted from our simulation, we focus primarily on the behavior of the electron density along the chain. This quantity is a key indicator of charge transport, as it allows us to distinguish between electron localization and delocalization.
In terms of the two-electron wave function, the density at location $x$ and time $t$ is given by
\begin{equation}
\begin{split}
 n_x(t) &= \langle \Psi(t) | n_x | \Psi(t) \rangle \\
&=  \sum_{x_1=1}^x  |\psi(x_1,x,t) |^2+ \sum_{x_2=x}^N  |\psi(x,x_2,t) |^2 
\end{split}
\end{equation}


\subsection{Defect-Mediated Pairing Mechanism: \\ \textit{When the Taxi Picks up its Passenger}}


In this first part of our study, we focus on the following situation. 
One electron  initially occupies the localized state at the defect, thus representing the passenger waiting for a taxi to arrive.
The second electron, representing the taxi driver, is described by a Gaussian wave packet propagating toward the defect. The initial two-electron wave function is thus defined as 
\begin{equation}
|\Psi(0) \rangle = \mathcal{C} \sum_{x_1=1}^N  e^{-(\frac{x_1-x_0}{\Delta x})^2} e^{ik_0 (x_1-x_0)}   |\Phi_{x_1,N}^\text{CSF} \rangle,
\end{equation}
where $\mathcal{C}$ is a normalization constant and we fixed $x_2=N$ with the wavepacket parameters $x_0=10$, $\Delta x=4$ and $k_0=1.3$. 
Along the same lines, it is interesting to note the existence of recent work that also focuses on the study of many-electron wavepackets in the context of the extended Hubbard Hamiltonian~\cite{al2013wave}.


\subsubsection{Time-evolution of the electronic density $n_N(t)$\\ on the site defect}

\begin{figure}[t]
    \centering
    \includegraphics[width=8.5cm]{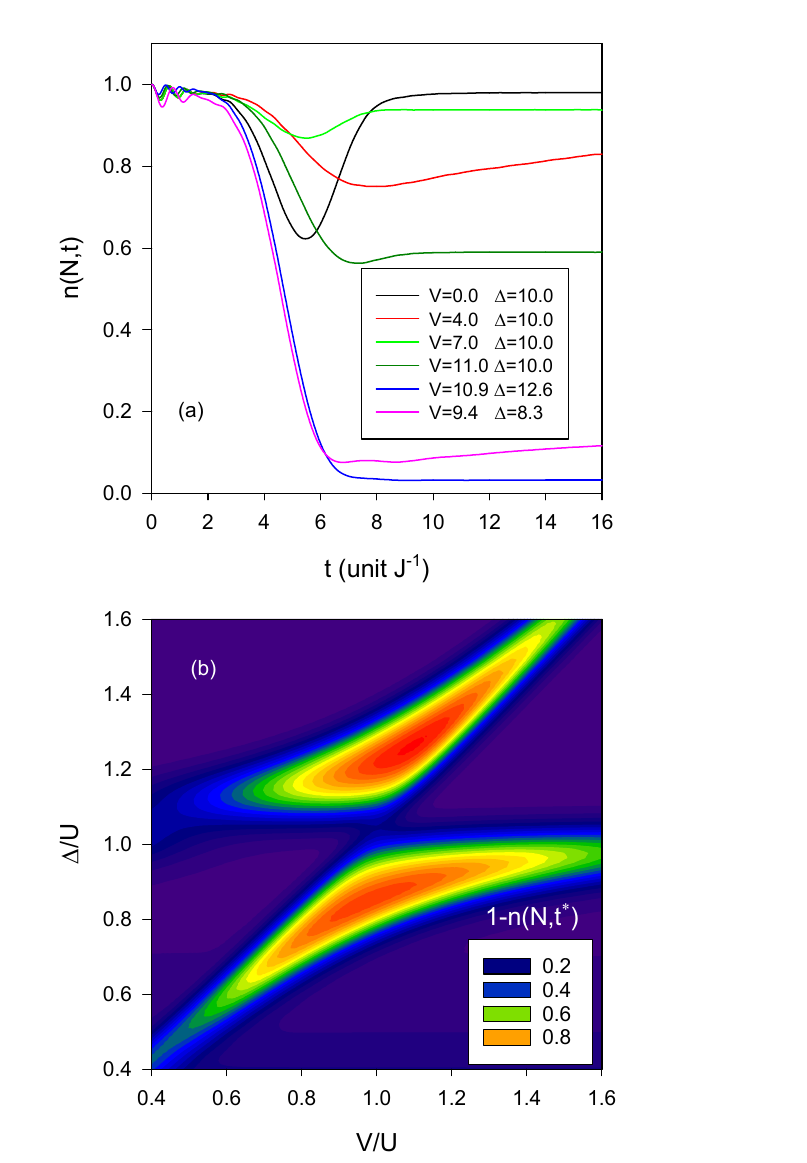}
    \caption{(a) Time evolution of the electron density $n_N(t)$ for $U=\Delta=10$ and for $V=0$ (black curve), $V=4$ (green curve), $V=7$ (red curve), and $V=11$ (yellow curve). Blue curve ($V=10.9$ and $\Delta=12.6$) and purple curve ($V=9.4$ and $\Delta=8.32$)  correspond to the time evolution when the parameters are optimized (see the text). (b) Variation of the QTE efficiency in the parameter space.
    } 
    \label{fig:taxi1}
\end{figure}

Let us first consider the time evolution of the electronic density $n_N(t)$ at the defect site. 
To this end, we assume the existence of a resonance between the TEBS and the SELS+SEFS by choosing $\Delta=10$.
Fig.~\ref{fig:taxi1}a shows the time evolution of the electronic density on the site defect $n_N(t)$ for $V=0$ (black curve), $V=4$ (green curve), $V=7$ (red curve), and $V=11$ (yellow curve). 
Initially equal to 1, the density $n_N(t)$ exhibits fast, low-amplitude oscillations at short times. 
It then decreases as $t$ approaches 4. From then on, the behavior depends strongly on the value of $V$.
For $V=0$, the density $n_N(t)$ reaches a minimum of 0.62 at $t=5.46$, before increasing to a value close to its initial value. At long times (on the time scale considered here) it shows only very small fluctuations around an average value of 0.98. In other words, although there is a resonance between the TEBS (which is unique at $V=0$) and the SELS+SEFS, there is no significant population transfer. 
The electron initially localized at the defect remains localized there with a probability of about $98\%$.
As $V$ increases, the behavior changes remarkably. \linebreak
While $n_N(t)$ still exhibits a minimum followed by an increase, its long-time value becomes significantly lower than one.
This effect is especially pronounced for $V=11$, where the density reaches a minimum of 0.56 at $t=7.38$, then increases and stabilizes around a nearly constant value of 0.59. 
\begin{figure*}
    \begin{center}
        \includegraphics[width=0.8\textwidth]{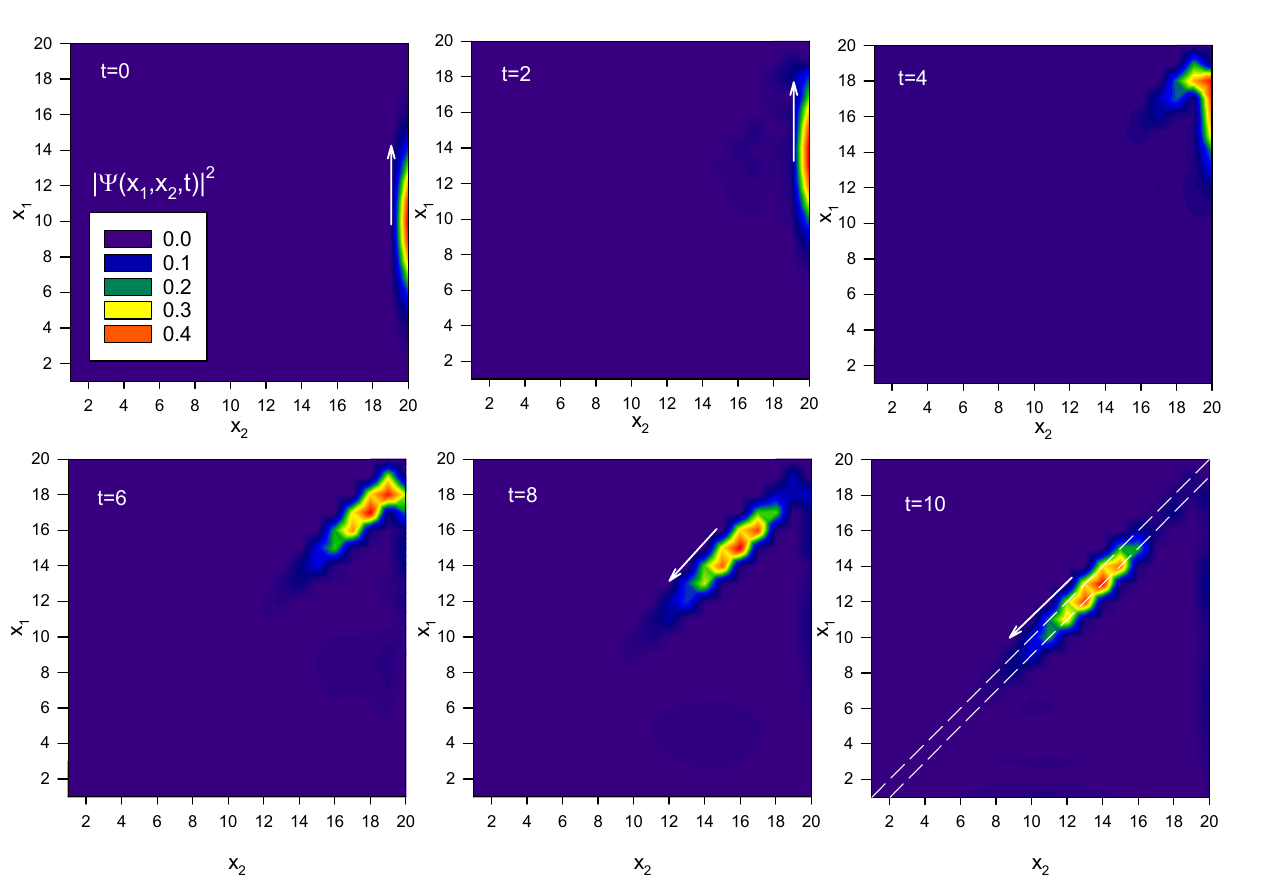}
    \end{center}
\caption{
{Time-evolution of the square modulus of the two-electron wave function $|\psi(x_1,x_2,t)|^2$ on the effective  2D-lattice.}
The different panels illustrate the delocalization of the probability density on the triangular connectivity graph associated with the singlet subspace, as shown in Fig.~\ref{fig:lattice}.
Here we consider the parameters $U=10$, $V=10.96$ and $\Delta=12.64$.
}
\label{fig:2DWF1}
\end{figure*}
The previous results show that the density at the defect strongly depends on the model parameters. In that context, we investigated the values taken by $n_N(t)$ after the ``interaction'' between the initial wave packet and the defect. To proceed, we fixed the time to $t^*=12$ and evaluated the density as a function of $V$ and $\Delta$. We introduced the parameter $\epsilon(t^*)=1-n(N,t^*)$, which characterizes the efficiency of the QTE as the amount of electronic density that has left the defect.
As illustrated in Fig.~\ref{fig:taxi1}b, there are two regions in the parameter space where $\epsilon(t^*)$ is optimized. First, when $V=10.96$ and $\Delta=12.64$, the efficiency reaches 0.97, indicating that $97\%$ of the electronic density has left the defect following the passage of the wave packet. Secondly, another region also leads to an optimization, although to a lesser extent. This occurs for $V=9.4$ and $\Delta=8.32$, parameter values for which the efficiency reaches $0.90$.

Based on the previous observations, purple and blue curves in Fig.~\ref{fig:taxi1}a illustrate the time evolution of $n_N(t)$ for parameter values that optimize $\epsilon(t^*)$. The blue curve corresponds to the optimal case with $V=10.96$ and $\Delta=12.64$, while the purple curve refers to the second optimum with $V=9.4$ and $\Delta=8.32$.
In both cases, following a short transient regime during which the density exhibits fast, low-amplitude oscillations near unity, $n_N(t)$ decreases rapidly between $t=3$ and $t=7$.
When the efficiency is maximized, the density drops from unity and tends toward a plateau at long times ($t>10$). In this case, it reaches a value approximately equal to 0.03, indicating that $97\%$ of the electronic population initially localized at the defect has left due to the interaction with the wave packet.
For the second set of optimal parameters, the behavior is slightly different. The density decreases from unity to reach a minimum of approximately 0.08 when $7<t<9$. Over this time interval, $92\%$ of the electronic population has left the defect. However, for $t>9$, the density increases slightly, gradually moving away from the optimal zone.

\subsubsection{Time-evolution of the effective single particle on the equivalent 2D-triangular Lattice}

To understand the underlying physics when the parameters are optimized, 
Fig.~\ref{fig:2DWF1} shows the time evolution of the square modulus  of the wave function $|\psi(x_1,x_2,t)|^2$ in the singlet subspace for $V=10.96$, and $\Delta=12.64$.
At time $t = 0$, $|\psi(x_1,x_2,t)|^2$ defines a Gaussian distribution with respect to the coordinate $x_1$, a distribution that is confined along the $x_2= N$ axis. It characterizes the initial state in which an electron is described by a Gaussian wave packet initially centered at $x_1 = x_0 = 10$, while the second electron occupies the defect ($x_2 = N$). At $t = 2$, this wave packet propagates along the $x_2 = N$ axis as time elapses, thus illustrating the motion of the first electron toward the defect. At times $t = 4$ and $t = 6$, the wave packet can be seen reflecting off the boundary of the singlet subspace, near the region where $x_1 = x_2 = N$.
Once this scattering process is complete, the reflected wave packet propagates along the parallel axes $x_2 = x_1$ and $x_2 = x_1 + 1$, as shown in Fig~\ref{fig:2DWF1} at $t=8$ and $t=10$. This indicates the formation of a bound state wave packet in which the two electrons remain confined close to each other and form a pair. 

To study the nature of the bound state wave packet, let us analyze the behavior of the two-electron wave function along the reaction pathway displayed in Fig.~\ref{fig:lattice}b. This pathway defines the active subspace that encompasses all resonant electronic configurations indexed by the integer $s$ ranging from 1 to $3N-4$. The odd values $s=1$, $3$, $...$, $2N-3$ refer to two electrons occupying the same site $x=1,...,N-1$. Even values $s=2$, $4$, $...$, $2N-2$ correspond to two electrons occupying two nearest-neighbor sites $x$ and $x+1$  with $x=1,...,N-1$. The values $s=2N-1$, $...$, $3N-4$ describe one electron trapped at the defect and one free electron occupying the sites $x=N-1,...,1$.

\begin{figure}
    \centering
    \includegraphics[width=7.5cm]{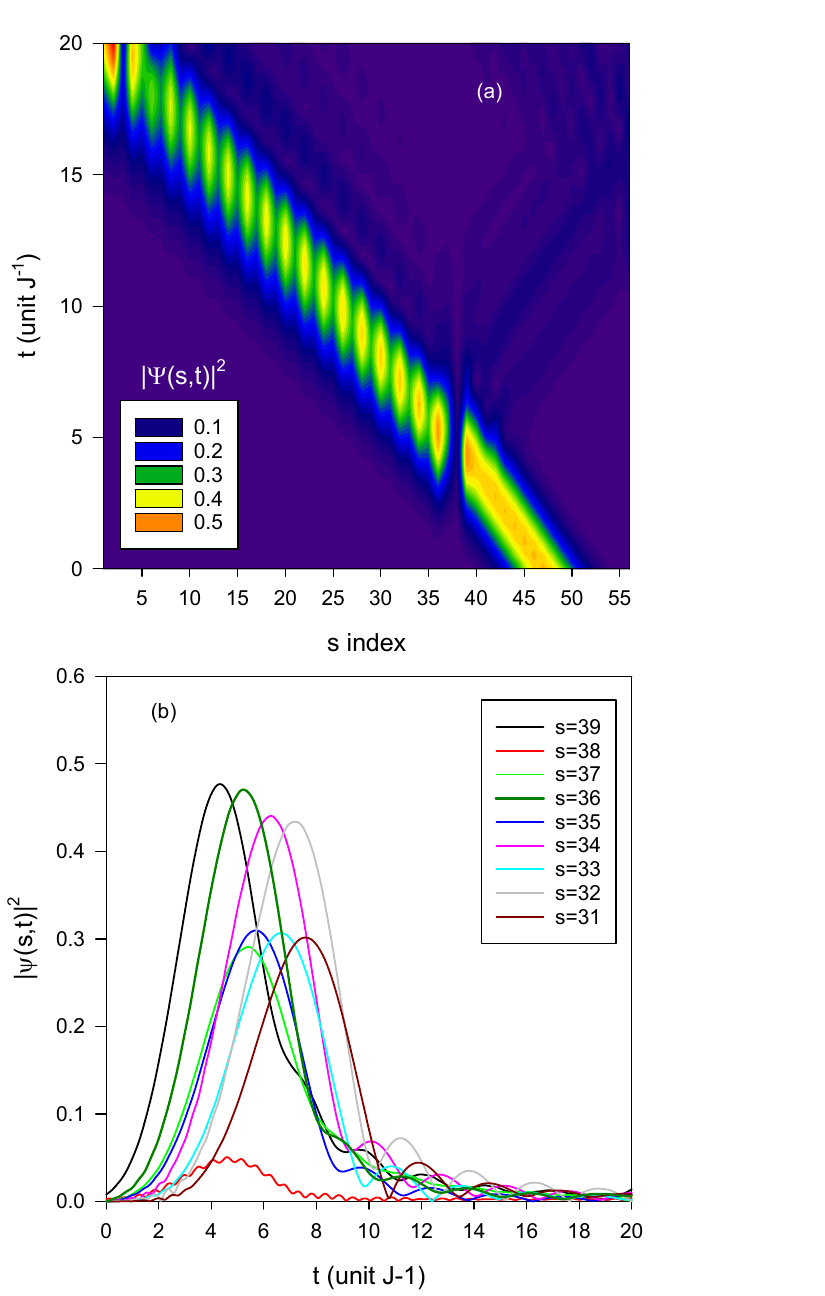}
    \caption{(a) Space-time evolution of the squared modulus of the two-electron wave function along the reaction pathway.
(b) Time evolution of the squared modulus of the wave function at selected points along the reaction pathway.}
    \label{fig:1DWF1}
\end{figure}

%
Fig.~\ref{fig:1DWF1}a clearly reveals two distinct propagation mechanisms. At short times, propagation occurs along the $x_2=N$ axis, where a ``continuous'' wave packet is observed along the corresponding segment of the pathway. Then, after the scattering process, the bound state wave packet that emerges exhibits a ``discontinuous'' structure along the reaction pathway. This discontinuity indicates that, in the transmitted wave packet, the weight of configurations involving two electrons on neighboring sites is greater than that of configurations where both electrons occupy the same site. This is a consequence of the internal structure of the bound states involved in the QTE.
This phenomenon becomes even more apparent in Fig.~\ref{fig:1DWF1}b, which shows the squared modulus of the wave function at selected points along the reaction pathway. First, at site $s=38$, which corresponds to one electron on site $N-1$ and the other on site $N$, the squared modulus of the wave function remains below 0.05. This indicates that this non-resonant configuration, whose energy is equal to $2\epsilon_0+V+\Delta$, plays only a minor role in the scattering of the initial wave packet. Next, at odd sites $s=$ 37, 35, 33, 31, the squared modulus is around 0.3. These sites correspond to configurations where both electrons occupy the same site $N-1,$ $N-2,$ $N-3$, and $N-4$. Finally, at the even sites $s=36$, $34$, $32$, the squared modulus exceeds 0.44. These sites correspond to configurations where the two electrons occupy neighboring lattice sites.
In other words, the bound state wave packet that propagates far from the defect mixes two configurations, namely, two electrons on the same site and two electrons on neighboring sites. But the configuration involving two electrons on neighboring sites has a larger weight to allow an almost perfect transmission at the interface between the TEBS and the SELS+SEFS. 

These results provide a scenario to interpret the QTE as follows. Since the localized electron occupies the $N$ site, a resonance occurs when the free electron arrives at site $N-2$. At this moment, the energy of this configuration is close to that of a pair of two electrons trapped on two neighboring sites. Therefore, the localized electron leaves the defect and jumps to the site $N-1$. It then forms a pair with the first electron which propagates away from the defect.
\begin{figure*} 
    \begin{center}
        \includegraphics[width=0.9\textwidth]{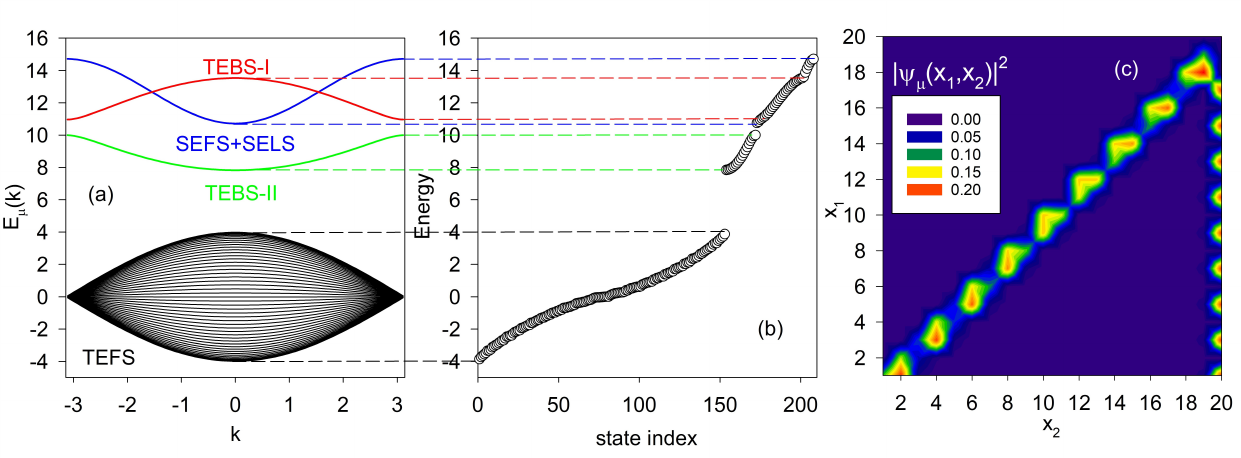}
    \end{center}
\caption{(a)  Two-electron energy spectrum of the extended Hubbard U-V model in an infinite translationally invariant lattice. (b) Two-electron energy spectrum of the finite-size chain with a defect ($N=20$). (c) Square modulus of the eigenstate 191 in the singlet subspace. The parameters are $U=10$,$V=10.96$ and $\Delta=12.64$.
}
\label{fig:spectrum1}
\end{figure*}
%


\subsubsection{Analysis of the two-electron energy spectrum}

To clarify the origin of this scenario, let us examine the two-electron energy spectrum displayed in Figs.~\ref{fig:spectrum1}a and ~\ref{fig:spectrum1}a for $V=10.96$ and $\Delta=12.64$. 
Fig.~\ref{fig:spectrum1}a shows the spectrum of the extended Hubbard model in an infinite translationally invariant lattice. In that case, the two-electron wave function being invariant with respect to a translation along the lattice, it is expanded as a Bloch wave whose wave vector $k$, which belongs to the first Brillouin zone of the lattice, is associated with the motion of the center of mass of the two electrons. Therefore, the spectrum shows the two-electron dispersion curves. It exhibits a quasi-continuum associated with the TEFS (black curves) centered on $2\epsilon_0$ with a bandwidth equal to $8J$. Above the quasi-continuum, two bands connected to two different bound states arise. The high-energy band refers to the TEBS-I (purple curve) whereas the low-energy band characterizes the TEBS-II (blue curve). Note that we add the dispersion curve of the state formed by one free electron in a Bloch state with wave vector $k$ and one electron localized at the defect (red curve). A clear resonance is then observed between this state  and the high-energy TEBS-I.

The energy spectrum of the finite-size chain with a defect is shown in Fig.~\ref{fig:spectrum1}b (note that the very high energies bound states TELBS-I and TELBS-II are not shown). The figure reveals that we recover the spectral properties of the infinite lattice. Indeed, by comparison, the states ranging 
from 1 to 153 define TEFS. By contrast, states ranging from 154 to 172 form the low-energy bound state band TEBS-II. This band extends from $7.83J$ to $9.99J$. A gap then appears between states $172$ and $173$. The states from 173 to 208 form a kind of continuum that includes both TEBS-I and SELS+SEFS.

The analysis of the eigenvectors reveals that states 173, 174, and 175 correspond to a ``free'' electron in an extended stationary state accompanied by an electron localized at the defect. The corresponding energies are $10.74J$, $10.82J$, and $10.94J$. This is consistent with the fact that in the infinite lattice, the SELS+SEFS band appears slightly below the TEBS-I band. The same properties are observed for states 203 to 208, whose energies lie between $13.85J$ and $14.69J$. These also characterize states in which a ``free'' electron in an extended stationary state is accompanied by an electron localized at the defect. In parallel with what happens in the infinite lattice, this phenomenon reflects the fact that the SELS+SEFS band appears slightly above the TEBS-I band.

The previous observations therefore suggest that the QTE is carried by the states corresponding to a superposition of the TEBS-I and SELS+SEFS bands. The relevant states are thus those between state 176 ($11.03J$) and state 202 ($13.59J$). The analysis of the eigenvectors reveals that the QTE is not the result of just a few eigenstates. In fact, it turns out that a significant number of eigenvectors are completely delocalized along the reaction pathway. As an example, Fig.~\ref{fig:spectrum1}c shows the eigenstate 191, with energy $12.70J$. This state corresponds to the one whose energy is closest to the crossing point between the TEBS-I and SELS+SEFS bands. The figure clearly shows the nearly perfect delocalization of the state along the reaction pathway. This delocalization enables ideal transfer between the two asymptotic ends of this pathway.

At this stage, it should be noted that, according to the band diagram in Fig.~\ref{fig:spectrum1}a, the crossing between the TEBS-I  band and the SELS+SEFS band occurs at $k=1.52$ and $E=12.685J$. When analyzing the eigenvectors of the extended Hubbard U-V model at this point, we observe that
the TEBS-I involved in the QTE at the resonance favors the configuration with two electrons bound at neighboring sites, whose corresponding weight is approximately equal to 63$\%$. However, the configuration with two electrons bound at the same site also contributes, but to a lesser extent, with a weight approximately equal to 36$\%$.

In conclusion, QTE can be interpreted as follows. The incident wave packet plays the role of the ``empty'' taxi driver. It moves toward the defect where the second electron is localized, acting as the waiting client. The client remains trapped on site $N$ until the taxi electron reaches site $N-2$. At that point, a resonance enables the client electron to move to site $N-1$, thereby forming a pair of two bound electrons on neighboring sites. Once this pair is formed, it propagates in the opposite direction through a mechanism that mixes two configurations, namely, two electrons on the same site and two electrons on neighboring sites. This feature reflects the idea that the taxi driver picks up the client and departs in the direction from which he came. 
The optimization of the QTE originates from the number of active eigenstates. Indeed, if the parameters are chosen such that a resonance occurs between the TEBS states and the SELS+SEFS states, then only a small number of states exhibit a significant delocalization along the reaction pathway. As a result, the QTE exists, but with a low probability. In contrast, when the parameters are optimized, the number of fully delocalized eigenstates along the reaction pathway becomes significantly larger, providing many channels for the wave packet transfer. 

Note that, as shown in Fig.~\ref{fig:taxi1}b, another region of the parameter space also leads to an enhancement of the QTE, but to a lesser extent. We have verified that the underlying physics is essentially the same. The main difference is that the QTE now arises from a resonance between the SELS+SEFS and the TEBS-II.

\subsection{Defect-Mediated Dissociation Mechanism: \\ \textit{When the Taxi Drops off its Passenger}}

In this second study, we focus on the following situation.
The two electrons are initially prepared in a superposition of bound states forming a pair wave packet localized far from the defect. This wave packet is constructed to propagate towards the defect so that the pair forms a quantum taxi in which one electron plays the role of the driver, while the other acts as the passenger.
To construct the wave packet, we proceed as follows.
Far from the defect, the quantum states of the system are isomorphic to those of an extended  Hubbard model in a translationally invariant lattice. As illustrated in Fig.~\ref{fig:spectrum1}a, the lattice supports two bound state bands lying above the TEFS continuum. In a bound state band $\mu$, the electrons form a pair whose center of mass $R=(x_1+x_2)/2$ delocalizes according to a Bloch wave with wave vector $k$. In contrast, the internal structure of the pair is described by a wave function $\psi_{\mu k}(m)$, which depends on the relative coordinate $m=x_2-x_1$ between the two electrons. This function is determined numerically. Under these conditions, a Gaussian wave packet relative to the bound state band $\mu$ in which the center of mass of the two electrons is initially localized around the position $R_0$ is defined as 
\begin{equation}
\ket{\Psi(0)} = \frac{ \mathcal{C}}{N} \sum_{x_1} \sum_{x_2} \sum_{k} e^{-\left(\frac{k-k_0}{\Delta k} \right)^2 } e^{ ik(R-R_0)} \psi_{\mu k}(m) |\Phi_{x_1,x_2}^\text{CSF} \rangle
\end{equation}
where $ \mathcal{C}$ is a normalization constant and $R_0=10$, $\Delta k=0.5$, and $k_0=-\pi/2$. Note that in the following of the text, we restrict our attention to
a wave packet formed from the TEBS-I band $\mu=1$. The analysis for the TEBS-II band $\mu=2$ is entirely analogous.

\begin{figure}
    \centering
    \includegraphics[width=7.5cm]{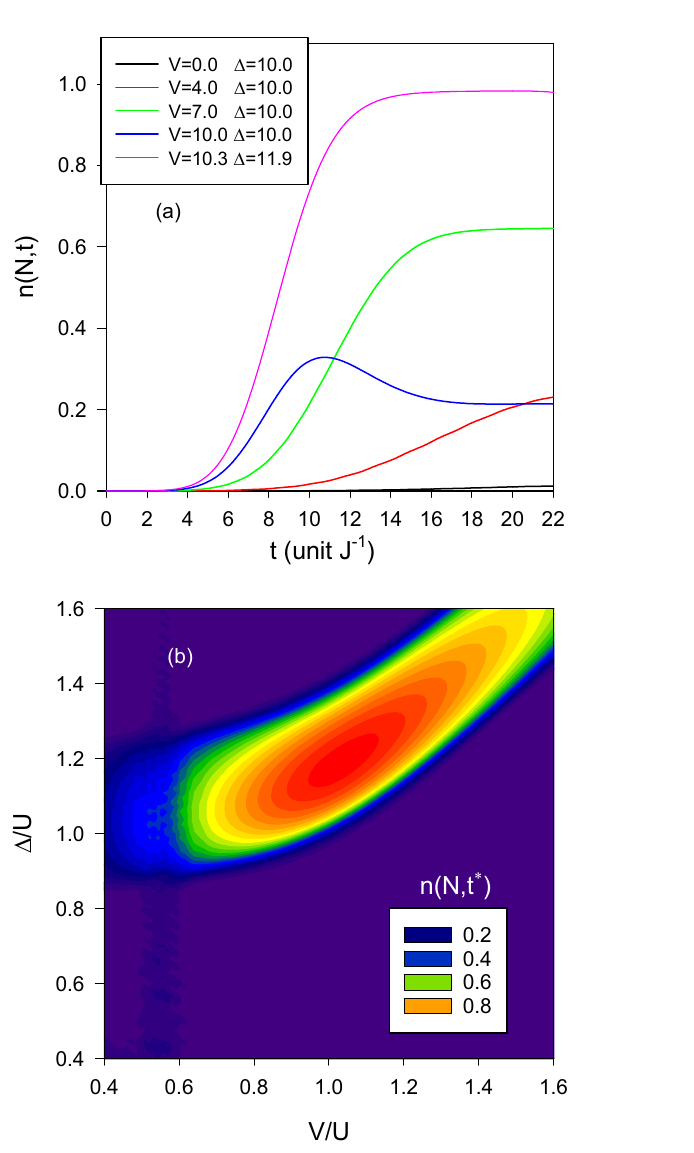}
    \caption{(a) Time evolution of the electron density $n_N(t)$ for $U=\Delta=10$ and for $V=0$ (black curve), $V=4$ (red curve), $V=7$ (green curve), and $V=10$ (blue curve). Purple curve ($V=10.36$ and $\Delta=11.92$)  corresponds to the time evolution when the parameters are optimized (see the text). (b) Variation of the QTE efficiency in the parameter space.
    } 
    \label{fig:taxi2}
\end{figure}
\begin{figure*} 
    \begin{center}
        \includegraphics[width=0.8\textwidth]{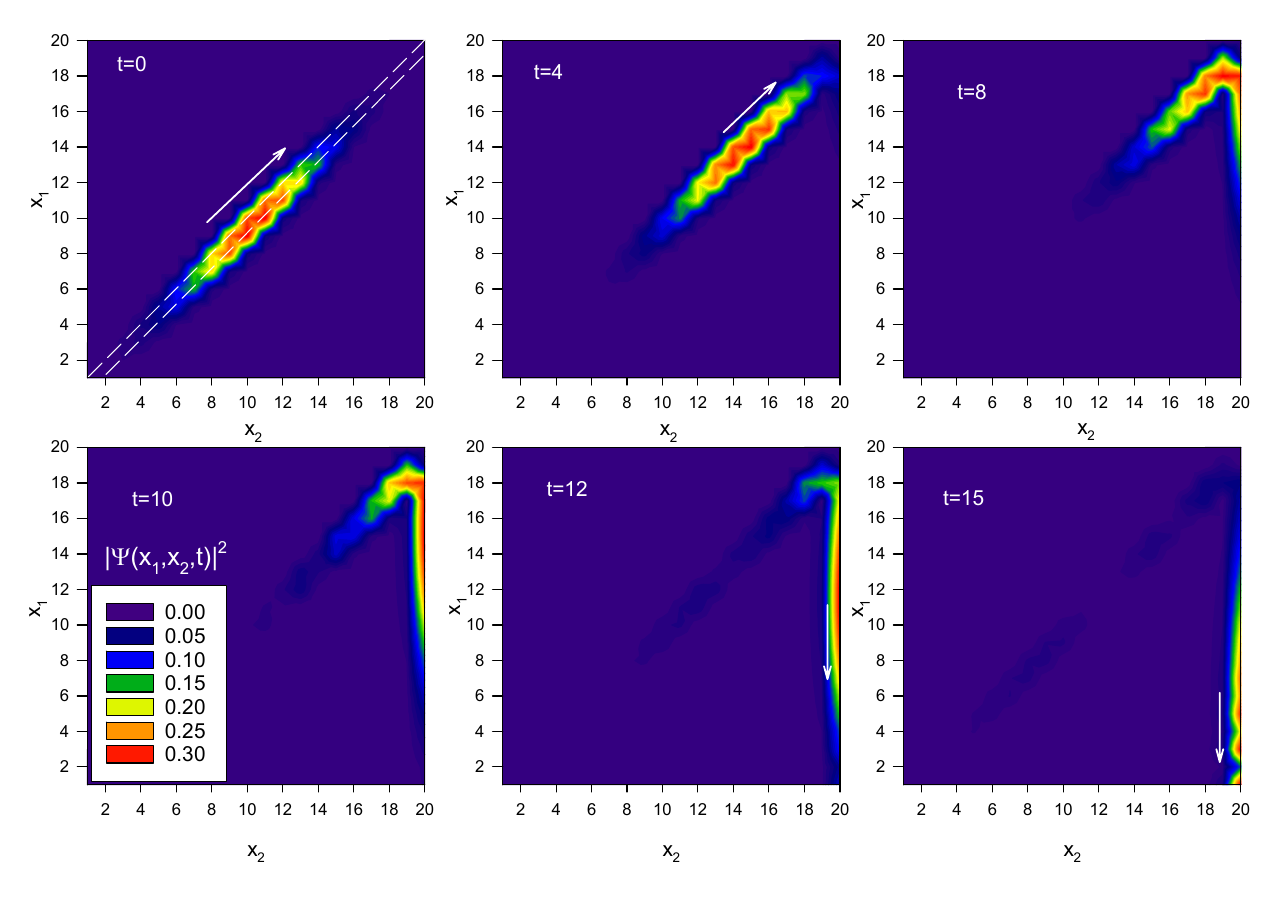}
    \end{center}
\caption{{Time-evolution of the square modulus of the two-electron wave function $|\psi(x_1,x_2,t)|^2$ on the effective  2D-lattice.}
The different pannel illustrate the delocalization of the probability density on the triangular connectivity graph associated with the singlet subspace, as shown in Fig.~\ref{fig:lattice}.
Here we consider the parameters for $U=10$,$V=10.36$ and $\Delta=11.92$.
}
\label{fig:2DWF2}
\end{figure*}

\subsubsection{Time-evolution of the electronic density $n_N(t)$\\ on the site defect}

The time evolution of the electronic density $n_N(t)$ at the defect site is shown in Fig.~\ref{fig:taxi2}a for $U=\Delta=10$ and for $V=0$ (black curve), $V=4$ (red curve), $V=7$ (green curve) and $V=10$ (blue curve). 
For $V=0$, initially equal to zero, $n_N(t)$ remains nearly zero at short times. It begins to increase around $t=5$ following a slowly varying straight line. 
Over the considered time scale, it reaches a maximum value equal to 0.013. In other words, although there is a resonance between the TEBS-I state (which is unique when $V=0$) and the SELS+SEFS state, there is no significant population transfer between the electron pair and the defect.
As $V$ increases, we see a different behavior. Indeed, at short times the density $n_N(t)$ remains small and starts to increase slowly from $t=5$. Beyond $t=5$, however, a more pronounced increase is observed, so that the density tends to a significant constant value over the time scale considered here. For $V=4$, $n_N(t)$ reaches 0.25 at $t=25$, which means that there is a one in four chance that the electron pair will dissociate, leaving one electron on the defect. This probability becomes even greater for $V=7$, reaching a value of approximately 64$\%$. For $V=10$ a peculiar behavior appears: the density reaches a maximum of 0.33 at $t=10.72$, then decreases, finally reaching 0.21 at $t=25$.

In accordance with the strategy employed in the previous study, the optimization of the parameters to maximize the density at the defect is achieved as follows. For $U=10$, we focus on the density value subsequent to the ``interaction'' between the initial wave packet and the defect. We fix $t^* = 20$ and calculate $n(N,t^*)$ as a function of $V$ and $\Delta$. As illustrated in Fig. \ref{fig:taxi2}b, there is a region in the parameter space where the population transfer between the pair wave packet and the defect is optimized. When $V=10.36$ and $\Delta=11.92$, $n(N,t^*)$ reaches a value of approximately 0.98.

Based on these observations, the purple curve in Fig.~\ref{fig:taxi2}a illustrates the time evolution of $n_N(t)$ when $V=10.36$ and $\Delta=11.92$. This evolution occurs according to three steps. At short times, typically between $t=0$ and $t=5$, the density is nearly zero, indicating that the pair wave packet has not yet reached the defect. Then, between $t=5$ and $t=12$, $n_N(t)$ increases almost linearly with time. Over this time interval, the pair wave packet interacts with the defect, leading to a population transfer from the pair to the defect. Finally, at long times ($t>16$), the electronic density on the defect reaches a constant, optimized value, equal to 0.98. The interaction between the pair wave packet and the defect is now complete. As a result, the pair dissociates, and an electron is trapped on the defect with a probability close to 100 $\%$.

\subsubsection{Time-evolution of the effective single particle on the equivalent 2D-triangular Lattice}

To illustrate the physical process that takes place when the parameters are optimized, the behavior of the wave function $|\psi(x_1,x_2,t)|^2$ in the singlet subspace is shown in Fig.~\ref{fig:2DWF2} at times $t=0, 4, 8, 10, 12$, and $15$ for $V=10.36$, and $\Delta=11.92$.
At $t = 0$, $|\psi(x_1,x_2,t)|^2$ defines a ``Gaussian'' distribution with respect to the center of mass coordinate of the two electrons.
It extends along the two lines $x_2=x_1$ and $x_2=x_1+1$, reflecting the fact that the initial bound state mainly contains configurations where the two electrons are either on the same site or on two nearest sites.
At $t = 4$ this wave packet propagates along the two lines $x_2=x_1$ and $x_2=x_1+1$ in the singlet subspace towards the site $x_1=x_2=N$. 
At $t = 8$ and $t = 10$ the reflection of the wave packet takes place at the boundary of the singlet subspace. Once this scattering process has taken place, the reflected wave packet moves back along the axis $x_2=N$.
This corresponds to the localization of one electron at the defect, while the second electron propagates away from the defect as a wave packet.
Note that one also observes a reflected component along the lines $x_2=x_1$ and $x_2=x_1+1$, indicating that a small part of the initial wave packet travels back in the opposite direction due to a reflection mechanism of the pair at the defect. However, the probability of this process remains small compared to the formation of a localized electron and a free electron.

To study the scattering process at the defect in detail, we have analyzed the nature of the two-electron wave function along the reaction path shown in Fig.\ref{fig:lattice}b. The following features were observed (not shown). First, for $s<38$, the occupation probability of the even sites along the reaction pathway is slightly higher than that of the odd sites. In the incident wave packet, the weight of configurations with two electrons on neighboring sites is more important than that of configurations with two electrons on the same site. 
Then, the squared modulus of the wave function at site $s=38$ remains below 0.05, indicating that the state with one electron at site $N-1$ and one electron at site $N$ does not play a significant role in the scattering process of the initial wave packet. 
Finally, the traversal of the wave packet at the interface between TEBS and SELS+SEFS is mediated by an effective transfer that occurs directly from site $s=36$ to site $s=39$.

To conclude this second study, it is worth mentioning that when the parameters are chosen to optimize the QTE, the two-electron energy spectrum and the corresponding eigenstates are very similar to those discussed in our first study.
Consequently, the QTE can be interpreted as follows. The initial wave packet, corresponding to a bound pair of electrons, describes the propagation of the pair towards the defect. Within this pair, one electron acts as the ``taxi driver'' while the other plays the role of the ``passenger'' trying to reach the defect site. As the pair propagates, it alternates between configurations in which both electrons occupy the same site and configurations in which they occupy adjacent sites. Upon approaching the defect, the pair first accesses the configuration where both electrons are at site $N-2$, followed by the configuration where one electron is at site $N-2$ and the other is at site $N-1$. At this critical point, a resonance condition allows the pair to dissociate: the taxi drops off the ``passenger'' electron, which moves from site $N-1$ to the defect site $N$, where it is trapped. Meanwhile, the ``taxi driver'' electron reverses direction and moves away from the defect.

\subsection{Both Mechanisms on a Chain with Two Defects: \\ \textit{When the Taxi Picks up and Drops off its Passenger} }

\begin{figure}
    \centering
    \includegraphics[width=7.5cm]{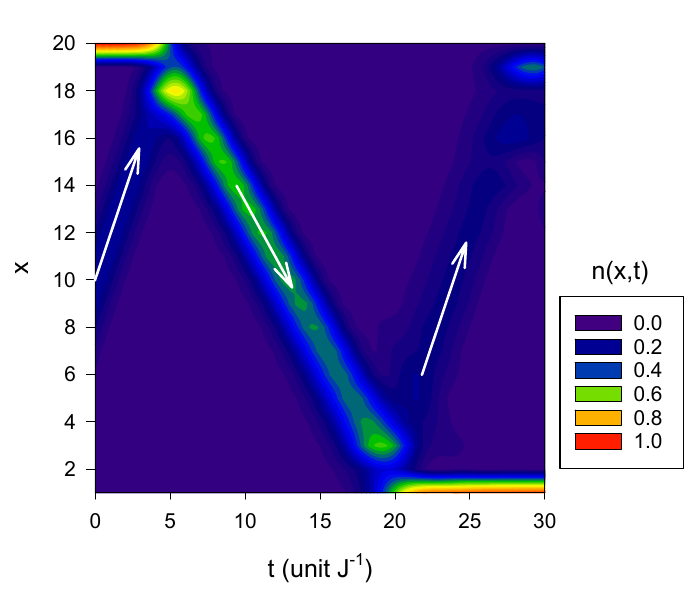}
       \caption{{Space-Time evolution of the electronic density $n(x,t)$ on the 1D Hubbard chain with two defects.} \linebreak
    Here we consider a 1D Hubbard chain whose size is fixed to $N=20$ carrying two defects on each extremity of the chain, i.e. $x=1$ and $x=N$, with same energy $\Delta=8.32$ (with   $V=9.4$ and $U =10$). }
    \label{fig:density}
\end{figure}

To conclude this numerical section, we present the spatio-temporal evolution of the electron density along the chain during the occurrence of the QTE. To this end, we slightly modify the system under study and consider a chain with two identical defects located at $x=1$ and $x=N$, respectively. This allows us to simultaneously observe the moments when the quantum taxi picks up and drops off its passenger. Therefore, at $t=0$, we assume that one electron is described by a Gaussian wave packet propagating toward the chain end $x=N$, whereas the second electron  occupies the localized state at the defect $x=N$. The parameters are fixed to optimize the QTE, i.e. $V = 10.9$ and $\Delta = 12.6$.

The space-time evolution of the electron density is shown in Fig.~\ref{fig:density}. This evolution clearly reflects the motion of a density wave between the two ends of the lattice. At short times, we observe the motion of the first electron, propagating as a single-particle wave packet toward the defect site. This wave packet travels with a velocity approximately equal to  $v=1.8J$. Over the same timescale, the electron density at the defect site $x=N$ remains close to unity, indicating that the second electron stays localized on its initial position. When the single-electron wave packet reaches the site $x=N-2$, a transfer of the electron population from the defect occurs, leading to its depopulation. This mechanism is accompanied by the emission of a new density wave packet, more intense than the first one, which propagates in the opposite direction. This second wave packet define a pair wave packet which travels more slowly, with a velocity approximately equal to $v=J$. It reaches the opposite end of the lattice ($x=1$) at a time of about $t=20$. At this time, when it reaches the site $x=3$, a density exchange is then observed between this wave packet and the defect located at $x=1$. The electron density of the defect increases to reach a value approximately equal to 0.95. This exchange process is accompanied by the emission of a third wave packet, isomorphic to the initial packet, which propagates towards the defect located at $x=N$ with a velocity close to $v=1.8J$.

In other words, Fig.~\ref{fig:density} highlights the phenomenon whereby one electron is manipulated by another. This is made possible by the existence of a resonance between bound states and localized states, which enables both the pairing of electrons on the one hand, and the dissociation of electron pairs on the other. Underlying this concept is the emerging idea that one electron enables the active transport of another electron between the two ends of a one-dimensional structure.

\section{Discussion}
 \label{sec:IV}
 
The previous results show that the extended  Hubbard model in a chain of finite size with a defect at the edge gives rise to a rich variety of two-electron quantum states, i.e. TEFS, TEBS, SELS+SEFS and TELBS. Depending on the region of the parameter space that is occupied, this richness allows us to observe remarkable dynamical processes when specific resonances arise between these different quantum states. In particular, we have identified the emergence of a novel dynamical process that we call the \textit{Quantum Taxi Effect} (QTE).

In order to observe this effect, several conditions have to be fulfilled. First, the model parameters must be such that a resonance exists between the SELS+SEFS band and a TEBS band. Second, the bound states involved in the process must have a specific internal structure. They must necessarily include the configuration where two electrons are trapped at two nearest neighbor sites. This in turn requires that the Coulomb interaction $V$ between two electrons occupying neighboring site be of the same order as the on-site Coulomb repulsion such as $V \sim U$. Under these conditions, the lattice exhibits two types of bound states, labeled TEBS-I (high-energy bound states) and TEBS-II (low-energy bound states), which appear above the continuum of the TEFS. As a result, there are two regions in the parameter space where the QTE is optimized. The existence of these two regions simply reflects the possibility of two types of resonances involving, on the one hand, the SELS+SEFS band and, on the other hand, either the TEBS-I band or the TEBS-II band.

At resonance, specific eigenstates emerge that are hybrid states which involve the superposition of SELS+SEFS and TEBS states. These hybrid states are delocalized within a region of the singlet subspace known as the reaction pathway. This pathway is formed by the defect lines that define the edges of the semi-infinite 2D lattice representing the Hubbard Hamiltonian in the singlet subspace. Consequently, the delocalization of the eigenstates along the reaction pathway allows the dissociation of a pair of electrons when it arrives at the defect, resulting in the localization of one electron on the defect while the other propagates away, far from the defect. Conversely, a free electron approaching the defect, which is already occupied by another electron, can form a pair with the localized electron by resonance, with the newly formed pair subsequently propagating away from the defect.

\begin{figure}[t]
    \centering
    \includegraphics[width=7cm]{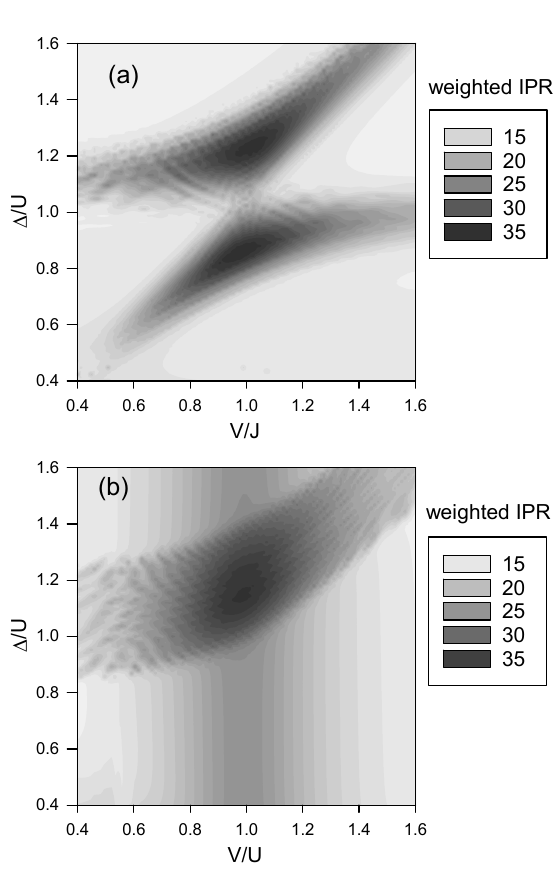}  
    \caption{{Weighted IPR in the ($U$,$V$)~parameter space.} \linebreak
     Here we consider $N=20$ and $U=10$, with the initial state that corresponds to \textbf{(a)} one electron localized at the defect, the second electron being described by a Gaussian wave packet propagating toward the defect, and \textbf{(b) }TEBS wave packet propagating toward the defect (see the text).} 
    \label{fig:IPR1}
\end{figure}

To interpret these observations, let us introduce a specific quantity named the weighted inverse participation ration $w\text{-IPR}$.
In terms of the two-electron eigenstates $\ket{\Psi_\mu}$, it is defined as  
\begin{equation}\label{eq:WIPR}
    w\text{-IPR} = \sum_{\mu} w_\mu \ \text{IPR}(\ket{\Psi_\mu}) 
\end{equation}
where the conventional inverse participation ratio (IPR) is expressed as 
 \begin{equation}
    \text{IPR}(\ket{\Psi_\mu})  =  \left( \sum_{x_1 x_2}|\bra{x_1,x_2 }\ket{\Psi_\mu}|^4\right)^{-1}
\end{equation}
In Eq.(\ref{eq:WIPR}), the normalized weight $w_\mu$ is defined in terms of the initial quantum state $\ket{\Psi(0)}$ considered in our numerical simulations (see Sec.~\ref{sec:IV}) such that  
\begin{equation}
    w_\mu = \frac{  |\bra{\Psi(0)}\ket{\Psi_\mu}|^2}{\sum_{\mu'} |\bra{\Psi(0)}\ket{\Psi_{\mu'}}|^2}
\end{equation}
As discussed in numerous papers, the IPR is a key quantity whose analysis allows us to distinguish between the localized or extended nature of states. Extended states have a very large IPR, whereas localized states are characterized by an IPR close to unity. By introducing the weighted IPR, we ensure that the most important eigenstates, which will contribute the most to the QTE, are the so-called ``active states'', i.e. those with the largest overlap with the initial two-electron quantum state $\ket{\Psi(0)}$. 
Given this, an eigenstate is considered an active state if and only if its overlap satisfies the threshold condition:
\begin{eqnarray}
   |\bra{\Psi(0)}\ket{\Psi_k}|^2  > \epsilon.
\end{eqnarray}
where $\epsilon$ is a small fixed threshold amplitude (set in our simulations to $\epsilon = 0.01 $).

Figs.~\ref{fig:IPR1} illustrates the behavior of the weighted IPR in the parameter space when $N=20$ and $U=10$. To proceed, we used the initial two-electron states defined in the previous section, i.e. 
one electron in wave packet propagating toward the defect accompanied by one electron localized at the defect (Fig.~\ref{fig:IPR1}a) and 
a TEBS-I wave packet (Fig.~\ref{fig:IPR1}b). In a perfect agreement with the numerical simulations, Fig.~\ref{fig:IPR1}a shows the occurrence of two domains in the parameter space where the weighted IPR exhibits significant values.  The first domain is defined by $9.1 <V<10.5$ and $8.2 <\Delta <8.9$ while the second domain is characterized by $9.60 <V<10.7$ and $12.0 <\Delta <12.8$. Within these two regions, the weighted IPR admits a maximum value approximately equal to 35. Note that, as mentioned previously, the occurrence of two domains results from the fact that the initial state belonging to the SELF+SEFS band, it can enter into resonance with either the TEBS-I band or the TEBS-II band. Conversely, as observed in Sec. III.B., Fig.~\ref{fig:IPR1}b shows the existence of a single region in the parameter space where the weighted IPR is maximized. In this case, it admits a maximum value, still approximately equal to $35$, within the domain defined by $8.70 <V< 11$ and $10.5 <\Delta < 13$. In that case, the initial state belonging to the TEBS-I band, it can present only a single resonance with the SELS+SEFS band.
\begin{figure}
    \centering
    \includegraphics[width=7.5cm]{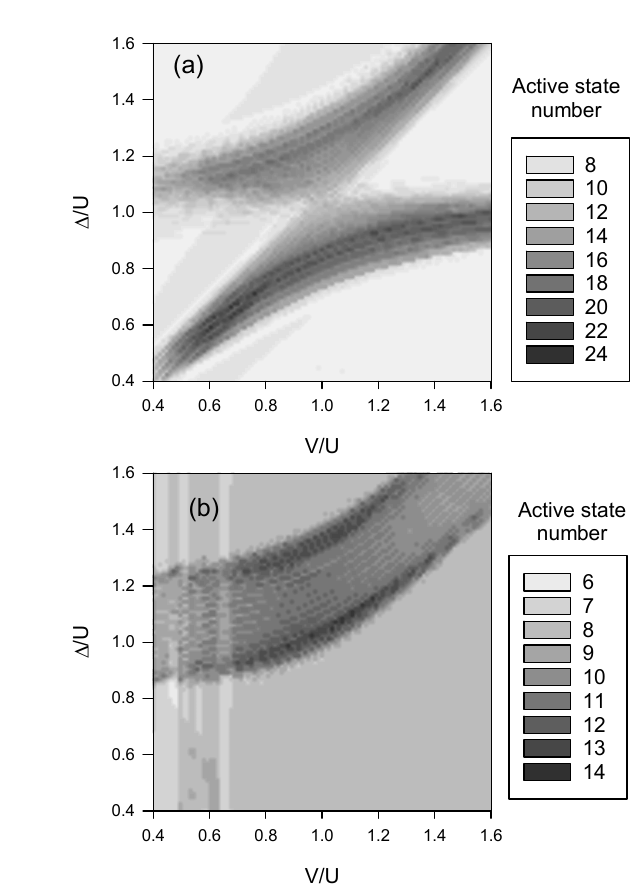}  
    \caption{Variation of number of active states in the parameter space for $N=20$ and $U=10$. The initial state corresponds to (a) one electron localized at the defect, the second electron being described by a Gaussian wave packet propagating toward the defect and (b) TEBS wave packet propagating toward the defect (see the text).} 
    \label{fig:IPR2}
\end{figure}
Whatever the initial state, the weighted IPR is of the order of 35. It thus represents $2/3$ of the length of the reaction pathway, which is made up of 56 sites when $N=20$ (see Fig.~\ref{fig:lattice}b). This illustrates perfectly the mechanism of delocalization of the active states along the reaction pathway when resonance occurs.

Nevertheless, the existence of a resonance is not the only condition required to observe an efficient QTE. Indeed, if the parameters are tuned to produce resonance, the number of significantly delocalized hybrid states along the reaction pathway is small a priori. In this case, the QTE takes place, but with a quite low probability. However, if the parameters are optimized, the number of perfectly delocalized hybrid states along the reaction pathway becomes large. This feature is illustrated in Fig.~\ref{fig:IPR2} which displays the behavior of the active state number in the parameter space for the two types of initial state. In regions of the parameter space where the weighted IPR is optimized, Fig.~\ref{fig:IPR2} clearly shows the sudden increases in the number of active states. Regardless of the initial conditions, the number of active states nearly triples depending on whether the parameters are optimized or not. In other words, the optimization of the parameters gives rise to the  occurrence of many channels allowing the transfer of the wave packet from one end  of the reaction pathway to the other. The probability of observing the QTE is enhanced, reaching a value close to 100$\%$.

\section{Conclusion}
 \label{sec:V}
 
In this work, we investigated the fundamental question of how local pairing and dissociation mechanisms between two strongly correlated electrons in a one-dimensional lattice may be triggered by the presence of a single site defect with amplitude $\Delta$.
To this end, we simulated the dynamics of a single electron pair on a chain with one defect at its extremity, using the extended Hubbard model that incorporates both the on-site local repulsion $U$ and the non-local inter-site repulsive Coulomb interaction $V$.
Restricting our attention to the singlet subspace, it has been shown that the two-electron problem can be mapped to the dynamics of a single fictitious particle moving on a 2D triangular lattice. The Coulomb repulsive interactions ($U$ and $V$), as well as the presence of the defect $\Delta$ in the real chain, introduce corresponding defects in the 2D lattice, leading to shifts in the effective site self-energies.
These defects distinguish between localized and delocalized states of the fictitious particle, which in turn correspond to localized, bound or free states of the two electrons. Therefore, the two-electron energy spectrum exhibits a rich variety of quantum states, i.e. TEFS, TEBS, SELS+SEFS and TELBS, that allows us to observe remarkable dynamical processes when specific resonances arise between these different quantum states. In particular, when $U\approx V \approx \Delta$, we have identified the emergence of a novel dynamical process, called the \textit{Quantum Taxi Effect} (QTE), which can be observed under certain conditions. 

First, the model parameters must allow for a resonance between the SELS+SEFS band and a TEBS band. Second, the relevant bound states must have an internal structure that includes configurations where two electrons occupy neighboring sites. This configuration is only possible when the non-local Coulomb $V$ is of the same order as the on-site Coulomb repulsion $U$. When these conditions are satisfied, the lattice supports two types of bound states, labeled TEBS-I (high-energy) and TEBS-II (low-energy), which lie above the TEFS continuum. These give rise to two distinct regions in parameter space where the QTE is optimized, corresponding to resonances with either TEBS-I or TEBS-II. At resonance, hybrid eigenstates form as superpositions of SELS+SEFS and TEBS states. These hybrid states are delocalized along a specific pathway within the singlet subspace, known as the reaction pathway. This delocalization enables two key processes: an incoming electron pair can dissociate at the defect, with one electron becoming localized and the other propagating away, or a free electron can approach a defect already hosting a localized electron, form a bound pair \textit{via} resonance, and then propagate as a new pair away from the defect. 

These findings demonstrate how the presence of a local defect can be harnessed to engineer nontrivial two-electron transport mechanisms in strongly correlated quantum systems, moving beyond conventional single-particle scattering and Anderson localization paradigms. They also highlight the critical role of internal pair structure and many-body spectral topology in enabling long-range, defect-assisted transport phenomena.

Looking ahead now, the conceptual framework introduced here with the \textit{Quantum Taxi Effect} may hold interesting potential for the development of emerging quantum technologies. 
For example, one potential application in quantum computing involves the controlled generation of entangled electron pairs in strongly correlated media for quantum information processing and communication.
In addition, this phenomenon could prove valuable in the study of out-of-equilibrium electronic structure problems with applications ranging from ultrafast dynamics to molecular electronics. 
Indeed, QTE may support new strategies for engineering molecular junctions with tailored electronic properties, paving the way for novel nano-devices (e.g. quantum sensors, photo-switches) at the crossroad of physics, chemistry and materials science.



\appendix 

\section{Equivalence between the 1D Hubbard chain and the 2D triangular lattice }
\label{app:Demo}

In this appendix, we will explain how to show the equivalence that occurs in the dynamics of the pair of electrons on the 1D-Hubbard chain and the one of a single effective particle evolving on the 2D triangular lattice as illustrated in Fig.~\ref{fig:lattice}.
To proceed, we start from the decomposition of the two-electron wavefunction $|\Psi(t) \rangle$ in the singlet CSF basis (see Eq.~(\ref{eq:basis})) which reads
%
\begin{equation}\label{eq:state}
|\Psi(t) \rangle = \sum_{x_1=1}^N \sum_{x_2=x_1}^N \psi(x_1,x_2,t) | \text{CSF}_{x_1,x_2} \rangle.
\end{equation}
%
We focus then on the time-dependent Schrodinger equation that govern the evolution of this wavefunction (\textit{via} the the singlet Configuration Hamiltonian given in Eq.~(\ref{eq:Ham_singlet})), 
\begin{equation}
    i \frac{d}{dt} |\Psi(t) \rangle = H^\text{Singlet} |\Psi(t) \rangle.
\end{equation}
Starting from this, we then project this equation on the different families of CSF states $| \text{CSF}_{x_1,x_2} \rangle$ to effectively exhibit how the 1D network will naturally connect to the 2D triangular network.

For example, when the two electrons do not occupy the same site and are also far away from the defect (\textit{i.e.}, when $x_1\neq x_2 \neq N$), the Schrodinger equation is expressed
as
\begin{equation}\label{eq:schrod1}
\begin{split}
   i\dot{\psi}(x_1,x_2,t)&=2\epsilon_0 \psi(x_1,x_2,t) \\
&-J[ \psi(x_1+1,x_2,t)+\psi(x_1-1,x_2,t) ]  \\
&-J [ \psi(x_1,x_2+1,t)+\psi(x_1,x_2-1,t) ]. 
\end{split}
\end{equation}
When the first electron is away from the defect ($x_1<N$) while the second one is located on the defect ($x_2=N$), the Schrodinger equation becomes
\begin{equation}\label{eq:schrod2}
\begin{split}
 i\dot{\psi}(x_1,N,t) & = [2\epsilon_0+\Delta] \psi(x_1,N,t) \\
 &- J\psi(x_1,N-1,t)   \\
&- J [ \psi(x_1+1,N,t) + \psi(x_1-1,N,t) ].
\end{split}.
\end{equation}
Similarly, when the second electron is located on the defect whereas the first electron occupies the nearest neighbor site ($x_1=N-1$ and $x_2=N$), the Schrodinger equation is written as
\begin{eqnarray}\label{eq:schrod3}
\begin{split}
i\dot{\psi}(N-1,N,t)&= [2\epsilon_0+V+\Delta] \psi(N-1,N,t)   \\
&-\sqrt{2}J\psi(N,N,t) \\
&-J \psi(N-2,N,t).
\end{split}
\end{eqnarray}
When the two electrons occupy the same site but away from the defect (\textit{i.e.} $x_1=x_2<N$), the Schrodinger equation is expressed as
\begin{equation}\label{eq:schrod4}
\begin{split}
 i\dot{\psi}(x_1,x_1,t)&= \left[ 2\epsilon_0+U \right] \psi(x_1,x_1,t)   \\
&- \sqrt{2}J \left[ \psi(x_1,x_1+1,t) + \psi(x_1-1,x_1,t) \right] 
\end{split}
\end{equation}
Similarly, when the two electrons are located on nearest neighboring sites  but remain away  from the defect (\textit{i.e.} $x_2=x_1+1<N$), the Schrodinger equation reads
\begin{equation}\label{eq:schrod5}
\begin{split}
 i\dot{\psi}(x_1,x_1+1,t)&= [2\epsilon_0+V] \psi(x_1,x_1+1,t)   \\
&-\sqrt{2}J[\psi(x_1,x_1,t)+\psi(x_1+1,x_1+1,t)]   \\
&-J[\psi(x_1-1,x_1+1,t)+\psi(x_1,x_1+2,t)]  
\end{split}
\end{equation}
Finally, when the two electrons occupy the defect site ($x_1=x_2=N$), the Schrodinger equation is written as
\begin{eqnarray}\label{eq:schrod6}
\begin{split}
i\dot{\psi}(N,N,t)&= [2\epsilon_0+U+2\Delta] \psi(N,N,t)   \\
&- \sqrt{2}J\psi(N-1,N,t)   
\end{split}
\end{eqnarray}
Note that all the other elements of the Schrodinger equation are obtained by symmetry due to the Hermitian nature of the Hamiltonian $H^\text{Singlet}$.

All the equations presented here (Eqs.(~\ref{eq:schrod1}-\ref{eq:schrod6})) clearly demonstrate the equivalence between the two-electron dynamics and the quantum mechanical dynamics of a single fictitious particle moving on the 2D lattice shown in Fig.~\ref{fig:lattice}. 
Within this framework, the two-electron wave function, $\psi(x_1,x_2,t)$, can be interpreted as the wave function of the fictitious particle. According to Eqs.~\ref{eq:schrod1}-\ref{eq:schrod6}, its dynamics is governed by a tight-binding Hamiltonian characterized by site self-energies and hopping constants coupling the nearest sites. The Coulomb repulsive interactions, as well as the presence of a defect in the real chain, introduce corresponding defects in the 2D lattice, leading to shifts in the self-energies. These defects distinguish between localized and delocalized states of the fictitious particle, which in turn correspond to localized, bound or free states of the two electrons.

Note that the similar equivalence was established two decades ago for two interacting boson-like excitations in a molecular chain. 
Special attention was given to the internal structure of the pair states as a function of the relative values of the one-site and lateral couplings between the bosons~\cite{pouthierPRE03,pouthierPRB05}.

\bibliography{mabiblio}

\end{document}